\documentclass[11pt,a4paper]{article}
\usepackage{jheppub}
\pdfoutput=1

\usepackage[utf8]{inputenc}
\usepackage{amssymb}
\usepackage{amsmath}
\usepackage{amsfonts}
\usepackage{graphicx}
\usepackage{xcolor}
\usepackage{xspace}
\usepackage[normalem]{ulem} 
\usepackage{lscape}
\usepackage{ wasysym }
\usepackage{float}
\usepackage{mathrsfs}
\usepackage{slashed}
\usepackage{multirow,rotating}

\usepackage{hyperref}
\hypersetup{
  colorlinks=true,
  allcolors=darkpurple,
  pdfborder={0 0 0},
  linktocpage=true
}

\definecolor{darkpurple}{rgb}{0.5,0,0.5}
\definecolor{cambridgeblue}{rgb}{0.64, 0.76, 0.68}
\definecolor{darkraspberry}{rgb}{0.53, 0.15, 0.34}

\def\gsim{\raise0.3ex\hbox{$\;>$\kern-0.75em\raise-1.1ex\hbox{$\sim\;$}}}
\def\lsim{\raise0.3ex\hbox{$\;<$\kern-0.75em\raise-1.1ex\hbox{$\sim\;$}}}

\newcommand{\ba}[1]{\begin{eqnarray} \label{(#1)}}
\newcommand{\ea}{\end{eqnarray}}

\def\gsim{\raise0.3ex\hbox{$\;>$\kern-0.75em\raise-1.1ex\hbox{$\sim\;$}}}
\def\lsim{\raise0.3ex\hbox{$\;<$\kern-0.75em\raise-1.1ex\hbox{$\sim\;$}}}

\newcommand{\calW}{{\cal W}}

\makeatletter
\g@addto@macro\bfseries{\boldmath}
\newcommand\Label[1]{&\refstepcounter{equation}(\mathrm{\theequation})\ltx@label{#1}&}
\makeatother

\graphicspath{{figs/}}

\allowdisplaybreaks

\preprint{}		

\title{Probing the dipole portal to heavy neutral leptons \\
via meson decays at the high-luminosity LHC}

\author[a]{Daniele Barducci,}
\emailAdd{daniele.barducci@df.unipi.it}
\affiliation[a]{Dipartimento di Fisica ``Enrico Fermi'', 
	Università di Pisa and INFN, Sezione di Pisa, \\
	Largo Bruno Pontecorvo 3, I--56127 Pisa, Italy}

\author[b]{Wei Liu,}
\emailAdd{wei.liu@njust.edu.cn}
\affiliation[b]{{\it Department of Applied Physics and MIIT Key Laboratory of Semiconductor Microstructure and Quantum Sensing, Nanjing University of Science and Technology, Nanjing 210094, China}}

\author[a]{Arsenii Titov,}
\emailAdd{arsenii.titov@df.unipi.it}

\author[c,d]{Zeren Simon Wang,}
\emailAdd{wzs@mx.nthu.edu.tw}
\affiliation[c]{Department of Physics, National Tsing Hua University, Hsinchu 300, Taiwan}
\affiliation[d]{Center for Theory and Computation, National Tsing Hua University, Hsinchu 300, Taiwan}

\author[e]{Yu Zhang}
\emailAdd{dayu@hfut.edu.cn}
\affiliation[e]{School of Physics, Hefei University of Technology, Hefei 230601, China}

\abstract{We consider the dipole portal to sterile neutrinos, also called heavy neutral leptons (HNLs).
The dipole interaction with the photon leads to HNL production in meson decays, as well as triggers the HNL decay into an active neutrino and a photon. 
HNLs with masses of order of 0.01--1~GeV are naturally long-lived if the dipole coupling is sufficiently small.
We perform Monte-Carlo simulations and derive the sensitivities of the proposed FASER2 and FACET long-lived particle experiments to HNLs produced via the dipole operator in meson decays at the high-luminosity LHC.
Our findings show that these future detectors will be complementary to each other, as well as to existing experiments, and will be able to probe new parts of the parameter space, especially in the case of the dipole operator coupled to the tau neutrino.
}

\begin{document}
\maketitle

%



\section{Introduction}\label{sec:intro}

The observation of neutrino oscillations has firmly established the non-vanishing mass of active neutrinos~\cite{Capozzi:2021fjo,Esteban:2020cvm,deSalas:2020pgw}, thus confirming the need for physics beyond the Standard Model (BSM).
While neutrino oscillations and cosmological measurements have placed bounds on the active neutrino masses (see \textit{e.g.}~Refs.~\cite{Canetti:2010aw,Planck:2018vyg,RoyChoudhury:2018gay,DiValentino:2021imh}), a large multitude of theoretical attempts have also been made to explain the experimental findings.
Among these theoretical efforts, a class of so-called ``seesaw'' models stand out, perhaps, as the most popular and elegant solution.
In particular, the type-I seesaw model~\cite{Minkowski:1977sc,Yanagida:1979as,Gell-Mann:1979vob,Mohapatra:1979ia,Schechter:1980gr} proposes the existence of three right-handed (RH) neutrinos $N_R$, which are gauge singlets under the Standard Model (SM) gauge group, and thus, usually referred to as sterile neutrinos. 
They possess Majorana masses and couple to the SM lepton doublets $L$ via a Yukawa interaction term $L\tilde{H}N_R$, with $H$ being the Higgs doublet ($\tilde{H} = i \sigma_2 H^\ast$).
Via the seesaw mechanism, the active neutrino masses are naturally suppressed by sterile neutrino masses of order $10^{16}$ GeV for $\mathcal{O}(1)$ Yukawa couplings.
The sterile neutrinos participate in the SM charged-current (CC) and neutral-current (NC) interactions, thanks to the mixing with the active neutrinos, allowing them to be directly tested in experiments.
Furthermore, they are highly motivated candidates for being also able to provide explanations for other SM shortcomings, such as the existence of baryon asymmetry and dark matter~\cite{Asaka:2005an,Asaka:2005pn,Boyarsky:2009ix}.

Although in the simplest type-I seesaw model the mixing parameters and the sterile neutrino masses are related, in more general incarnations of the seesaw mechanism~\cite{Wyler:1982dd,Mohapatra:1986bd,Bernabeu:1987gr,Akhmedov:1995ip,Akhmedov:1995vm,Malinsky:2005bi} they can be independent parameters, thus allowing for richer phenomenology at colliders and other experiments.
For instance, in recent years sterile neutrinos with masses below the electroweak (EW) scale, more often called heavy neutral leptons (HNLs) in the context of collider searches, have received increasingly more attention, see \textit{e.g.} Refs.~\cite{Bondarenko:2018ptm, Bryman:2019bjg, Abdullahi:2022jlv, Deppisch:2018eth,Deppisch:2019kvs, Liu:2021akf,  Liu:2022kid, Liu:2022ugx, Liu:2023gpt}.
Such HNLs can be produced, via active-sterile neutrino mixing, from rare decays of the Higgs boson, $W$- and $Z$-bosons, as well as QCD pseudoscalar and vector mesons.
These different production modes give rise to different sensitivities on the HNL masses.
So far, ATLAS and CMS have performed searches for HNLs with mass $\sim 3-30$ GeV with leptonic final states, mainly produced in $W$-boson decays and in association with a prompt charged lepton~\cite{ATLAS:2022atq,CMS:2022fut}.

Beyond the standard active-sterile mixing,  HNLs could also interact with SM particles via effective non-renormalizable dipole  operators~\cite{Magill:2018jla} suppressed by powers of a cut-off scale, which parameterizes the scale of breakdown of the effective description.
These operators couple the HNLs to SM EW gauge bosons and can, in general, enhance the production and decay rates of the HNLs.
For instance, after EW symmetry breaking the CC dipole operator coupling the $W$-boson, a charged lepton, and an HNL, can induce leptonic and semi-leptonic meson decays into an HNL, if the process is kinematically allowed, even though this process will turn out to be highly suppressed by the mass of the $W$-boson.
On the other side, the NC dipole operator with a photon can result in both a new HNL production mode via meson decays and also a new decay mode for the HNL, namely $N\to \nu\gamma$. 
This can be the dominant HNL decay channel, if one assumes a negligible contribution from the active-sterile mixing, which is the case for HNL masses $m_N\lesssim 10$ GeV and cut-off scales below  $\sim 100\;$TeV~\cite{Barducci:2022hll,Liu:2023nxi}.
This dipole operator with a SM photon will be the main subject of our study.

Various phenomenological studies have been performed for sterile neutrinos and dipole operators at different experimental setups~\cite{Aparici:2009fh, Giunti:2014ixa, Aparici:2013xga, Coloma:2017ppo, Abazajian:2017tcc,Shoemaker:2018vii, Magill:2018jla,  Butterworth:2019iff, Chala:2020vqp,   Chala:2020pbn, Brdar:2020quo, Barducci:2020icf, Plestid:2020vqf, Jodlowski:2020vhr,  Balaji:2019fxd, Balaji:2020oig, Schwetz:2020xra, Ismail:2021dyp,Miranda:2021kre,Dasgupta:2021fpn,Atkinson:2021rnp, Kamp:2022bpt,Gustafson:2022rsz,Huang:2022pce,Li:2022bqr,Acero:2022wqg,Feng:2022inv,Hati:2022tfo,Mathur:2021trm,Bolton:2021pey,Ovchynnikov:2022rqj,Zhang:2022spf, Zhang:2023nxy,   Ovchynnikov:2023wgg,Guo:2023bpo,Fernandez-Martinez:2023phj,  Delgado:2022fea, Duarte:2020vgj}.%
\footnote{For early works on electromagnetic properties of neutrinos see Refs.~\cite{Petcov:1976ff,Pal:1981rm,Shrock:1982sc}.}
Here we elaborate on a few of them.
For example, Ref.~\cite{Jodlowski:2020vhr} considered NC dipole operators for the neutrino physics program experiment FASER$\nu$~\cite{FASER:2019dxq,FASER:2020gpr} at the LHC, and Ref.~\cite{Barducci:2022gdv} focused on a dipole operator with two HNLs with masses at the GeV-scale and worked out the sensitivities of a series of future LHC facilities such as FASER2~\cite{Feng:2017uoz,FASER:2018eoc}, MoEDAL-MAPP~\cite{Pinfold:2019nqj,Pinfold:2019zwp}, and SHiP~\cite{SHiP:2015vad,SHiP:2018yqc}.
Furthermore, Refs.~\cite{Masip:2012ke,Fischer:2019fbw} showed that a sub-GeV sterile neutrino decaying to $\nu\gamma$ could accommodate the neutrino excess at MiniBooNE, and Refs.~\cite{Cheung:2022luv,Smirnov:2022suv,Brdar:2022rhc} argued for a sub-MeV sterile neutrino with the magnetic dipole operator being able to explain the high-energy gamma ray bursts observed at LHAASSO~\cite{LHAASO}.
The same signature was also studied in Refs.~\cite{Kling:2022ehv,Dreiner:2022swd,Dienes:2023uve} for FASER and FASER2, however in the framework of axion-like particles, light binos in R-parity-violating supersymmetry, and inelastic-dipole dark matter, respectively.
Further, Ref.~\cite{Jodlowski:2023fvz} investigated the sensitivity of FASER2 to a long-lived light bino as the next-to-lightest supersymmetric particle decaying into a single photon and the lightest supersymmetric particle (ALPino or gravitino).
Finally,  some of us recently estimated in Ref.~\cite{Liu:2023nxi} the sensitivities of similar proposed far-detector programs at the LHC to HNL dipole operators related to the EW gauge bosons, assuming the HNLs are produced directly in $pp$ collisions.

In this work, we will consider HNLs\footnote{We will work under the simplified assumption that only one HNL is kinematically accessible.} produced from the decays of QCD mesons, which in turn are abundantly produced at the various LHC interaction points (IPs).
The meson decays are triggered by the NC dipole operator with a photon, and the same operator is also responsible for the HNL decay into a neutrino and a photon.
If the dipole coupling responsible for the HNL decay is sufficiently suppressed, this particle is long-lived and thus travels largely in the forward direction, flying towards far-detector experiments such as FASER(2) and FACET~\cite{Cerci:2021nlb}.\footnote{A whole class of far detectors have been proposed to be operated in the vicinity of the different IPs at the LHC, primarily aimed to search for long-lived particles,
see Refs.~\cite{Alimena:2019zri,Lee:2018pag,Curtin:2018mvb,Knapen:2022afb} for recent reviews.
In this study, we only include the concepts for which technical details are already given regarding the detection of a single photon.}
The HNLs can then potentially decay into a neutrino and a highly energetic photon inside these detectors, where this displaced photon can be observed via electromagnetic showers.
It is the purpose of this paper to investigate the exclusion limits on the sterile neutrino mass and dipole operator scale that can be obtained at these experiments.

The paper is organized as follows.
In the next section, we discuss the sterile neutrino dipole operators and compute the corresponding production and decay rates for the HNL.
Then in Sec.~\ref{sec:exp}, we introduce and describe the experiments of interest, discuss possible background sources, and explain the Monte-Carlo simulation and computation procedures that we have used for estimating the sensitivities.
We then present our numerical findings in Sec.~\ref{sec:results}, before concluding the paper with a summary in Sec.~\ref{sec:conclusions}.


\section{Theoretical framework}\label{sec:model}
In this section, we describe the theoretical framework that we employ throughout our analysis.
We extend the SM with a Dirac sterile neutrino $N$, which is a total singlet under the SM gauge group.
Below the EW scale, we parameterize the low-energy dipole portal with the photon and the left-handed active neutrino $\nu_L$ through a $d=5$ operator as\footnote{We assume the presence of an additional symmetry that prevents from writing a similar operator with $N_L^c$, where the superscript $c$ indicates the charge-conjugate field.}
\begin{equation}
    \mathcal{L} = d_\gamma^\alpha\, \overline{\nu_{\alpha L}} \sigma^{\mu\nu} N_R F_{\mu\nu} + \text{h.c.} \, ,
    \label{eq:dipole_IR}
\end{equation}
where $\alpha = e,\mu,\tau$ is a flavor index, $2\sigma_{\mu\nu}= i [\gamma_\mu,\gamma_\nu]$, and $F_{\mu\nu}$ is the electromagnetic field strength tensor.
Above the EW scale, this interaction must be described by $d=6$ operators that respect the full ${\rm SU}(2)_L \otimes {\rm U}(1)_Y$ SM gauge symmetry.
They can be parameterized as~\cite{Magill:2018jla}
\begin{equation}
	\mathcal{L} = \overline{L_\alpha} \left(d_{\calW}^\alpha\, W_{\mu \nu}^a \tau^a + d_{\cal B}^\alpha\, B_{\mu \nu}\right) \tilde{H}\sigma^{\mu \nu} N_R + \text{h.c.} \, ,
	\label{eq:LWB}
\end{equation}
where $\tilde{H}=i\sigma_2H^*$, $\tau^a=\sigma^a/2$ with $\sigma^a$ being Pauli matrices, $W_{\mu \nu}^a$ and $B_{\mu \nu}$ denote the ${\rm SU}(2)_L$ and ${\rm U(1)}_Y$ field strength tensors with $W_{\mu \nu}^a\equiv\partial_\mu W_\nu^a-\partial_\nu W_\mu^a + g \varepsilon^{abc} W_\mu^b W_\nu^c$ and $B_{\mu\nu}\equiv\partial_\mu B_\nu-\partial_\nu B_\mu$, and $L_\alpha$ are the SM lepton doublets.
Once the Higgs boson acquires its vacuum expectation value $\langle H \rangle = v/\sqrt 2$, one obtains
\begin{equation}
	\mathcal{L} \supset d_W^\alpha\, \overline{\ell_{\alpha L}} W^-_{\mu \nu} \sigma^{\mu \nu} N_R + \overline{\nu_{\alpha L}}\left(d_\gamma^\alpha\, F_{\mu \nu}+d_Z^\alpha\, Z_{\mu \nu}\right) \sigma^{\mu \nu} N_R + \text{h.c.} \, ,
	\label{eq:LWZk}
\end{equation}
with\footnote{Note that $d_{\gamma,Z,W}^\alpha$ have a different mass dimensions with respect to $d_{\calW,\cal{B}}^\alpha$.}
\begin{equation}
\begin{aligned}
	d_\gamma^\alpha&=\frac{v}{\sqrt{2}}\left(d_{\cal B}^\alpha\cos\theta_{w}+\frac{d_{\calW}^\alpha}{2}\sin\theta_w\right)\,,\\ 
	d_Z^\alpha&=\frac{v}{\sqrt{2}}\left(\frac{d_{\calW}^\alpha}{2}\cos\theta_{w}-d_{\cal B}^\alpha\sin\theta_w\right)\,,\\
	d_W^\alpha&=\frac{v}{2}d_{\calW}^\alpha \, ,
\end{aligned}
\label{eq:broken_unbroken_relation} 
\end{equation}
where $\theta_w$ is the weak mixing angle.
In writing Eq.~\eqref{eq:LWZk} we have neglected interactions arising from the non-Abelian structure of the ${\rm SU}(2)_L$ interactions.
The operators in Eq.~\eqref{eq:LWB} thus induce dipole operators to SM photons as well as to the weak gauge bosons $Z$ and $W$ with the couplings $d_\gamma^\alpha$, $d_Z^\alpha$, and $d_W^\alpha$.
For a given lepton flavor, the dipole couplings $d_\gamma^\alpha$, $d_Z^\alpha$, and $d_W^\alpha$ in the broken phase are linearly dependent, and determined by only two parameters $d_{\calW}^\alpha$ and $d_B^\alpha$ in the unbroken phase via Eq.~\eqref{eq:broken_unbroken_relation}.
Overall, for a fixed lepton flavor $\alpha$, the model is thus described by the three parameters
\begin{equation}\label{eq:freepara1}
	\{m_N, d_{\calW}^\alpha, d_{\cal B}^\alpha\} \ ,
\end{equation}
where $m_N$ is the Dirac mass of the sterile neutrino.
Further, if we assume the relation $d_{\calW}^\alpha = a \times d_{\cal B}^\alpha$, we have
\begin{equation}
\begin{aligned}
	d_Z^\alpha &=\frac{d_\gamma^\alpha (a\cos\theta_w-2\sin\theta_w)}{2\cos\theta_w+a\sin\theta_w}\,,\\
	d_W^\alpha &= \frac{\sqrt{2}a d_\gamma^\alpha}{2\cos\theta_w+a\sin\theta_w} \, ,
\end{aligned}\label{eq:dzdw}
\end{equation}
and the independent parameters of Eq.~\eqref{eq:freepara1} can be replaced by the parameters
\begin{equation}
	\{m_N, d_\gamma^\alpha, a\} \ .
\end{equation}
As we will show, because of the large mass difference that exists between the $W,Z$-bosons and the QCD bound states, the contributions to their decay widths arising from the $d_{W,Z}^\alpha$ dipole operators are extremely suppressed for ranges of parameters still allowed by current experimental constraints.
For this reason, in Sec.~\ref{sec:results} we will present our results focusing only on the $d_\gamma^\alpha$ dipole operator with the photon at $d=5$ given by Eq.~\eqref{eq:dipole_IR}, bearing however in mind that its origin resides in $d=6$ operators,\footnote{One could also write a dipole interaction among sterile states $\overline{N_L}\sigma^{\mu\nu}N_R B_{\mu\nu}$ that gives rise to an active-sterile dipole portal as the one of Eq.~\eqref{eq:dipole_IR} after a mixing insertion. 
Even though this could be an interesting scenario to analyze, the arising effects are expected to be extremely suppressed by the mixing angle.
We thus decide not to inspect this scenario further.} as described in Eq.~\eqref{eq:LWB}.

\subsection{HNL production in meson decays}

\begin{figure}[t]
 \centering
 \includegraphics[width=\textwidth]{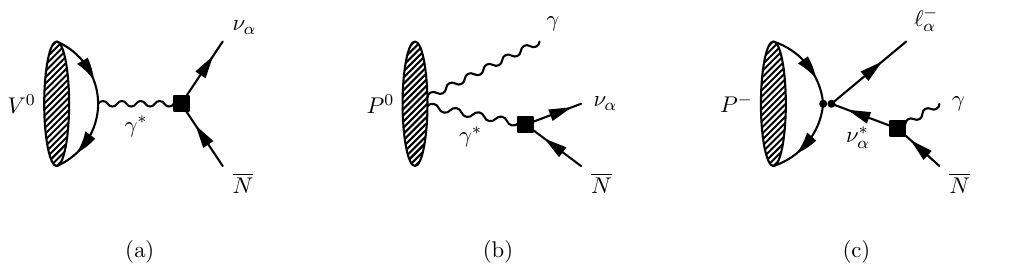}
 \caption{Two-body (a) and three-body (b),  (c) meson decays triggered by the neutrino dipole operator $d_\gamma^\alpha$, which is 
 represented by a solid square.}
 \label{fig:decays}
\end{figure}

We now discuss the decays of QCD bound states induced by the dipole operators previously introduced.
The photon dipole triggers the two-body decays $M \to \nu_\alpha \overline{N}$, with $M$ denoting a meson, see Fig.~\ref{fig:decays}a.
In the case of a pseudoscalar meson, $M=P$, the corresponding amplitude vanishes because
\begin{equation}
 \langle 0 | \overline{q_i} \gamma^\mu q_j | P(p) \rangle = 0\,. 
\end{equation}
For a vector meson, $M=V$, one has
\begin{equation}
 \langle 0 | \overline{q_i} \gamma^\mu q_j | V(p,\epsilon) \rangle = i f_V m_V \epsilon^\mu\,, 
\end{equation}
where $m_V$ and $f_V$ are the mass and the decay constant of $V$, respectively, and $\epsilon^\mu$ is its polarization vector. 
We collect the numerical values of the meson masses and decay constants in App.~\ref{sec:appendix}.
The decay width of interest reads
\begin{equation}
 \Gamma\left(V^0 \to \nu_\alpha \overline{N}\right) = \frac{f_V^2 m_V}{12\pi} 
 e^2 Q_q^2 \left|d_\gamma^\alpha\right|^2  
 \left(1-\frac{m_N^2}{m_V^2}\right) \left(1+\frac{m_N^2}{m_V^2}-2\frac{m_N^4}{m_V^4}\right)\,,
 \label{eq:GammaV0nuNDir}
\end{equation}
where 
$e$ is the electron charge, and $Q_q$ is the charge (in units of $e$) of quark $q$ composing $V^0 \sim q \overline{q}$. 
We focus on flavorless mesons $V^0 = \rho^0$, $\omega$, $\phi$, $J/\psi$, $\Upsilon(1S)$, which decay at tree level.
For $V^0 = \rho^0$, the charge $Q_q \to (Q_u-Q_d)/\sqrt{2} = 1/\sqrt{2}$, 
and for $V^0 = \omega$, $Q_q \to (Q_u+Q_d)/\sqrt{2} = 1/(3\sqrt{2})$.

The dipole operator $d_W^\alpha$ induces the two-body decays of charged mesons $M^\pm \to \ell_\alpha^\pm N$, mediated by an off-shell $W^\pm$ boson.
For a pseudoscalar meson, $M^\pm = P^\pm$, the amplitude for such process scales as $p_\mu p_\nu \sigma^{\mu\nu}$ ($p$ being the meson four-momentum), and hence vanishes, \textit{cf}.~Ref.~\cite{Alcaide:2019pnf}.
For a vector meson, $M^\pm = V^\pm$, the corresponding rate reads $\Gamma(V^\pm \to \ell_\alpha^\pm N) \propto |d_W^\alpha|^2 f_V^2 m_V (m_V/m_W)^4$ and is thus strongly suppressed by a large power of the ratio of the vector meson mass over the $W$-mass.
Similar considerations apply to the operator $d_Z^\alpha$.

In addition to the two-body decays of flavorless vector mesons, we have computed three-body decays of pseudoscalar mesons, see also Ref.~\cite{Ovchynnikov:2022rqj}.
The first class of three-body decays is depicted in Fig.~\ref{fig:decays}b, where a neutral pseudoscalar meson $P^0 = \pi^0$, $\eta$ or $\eta'$, decays to two photons via the chiral anomaly, with one of the photons subsequently converting to $\nu_\alpha$ and $\overline{N}$ via $d_\gamma^\alpha$.
To compute the corresponding decay rates, we parameterize the $P^0 \to 2\gamma$ decay amplitude as~\cite{Peskin:1995ev}
\begin{equation}
 i \mathcal{M}\left(P^0 \to 2\gamma\right) = i \frac{e^2}{4\pi^2} \frac{1}{f_P} 
 \epsilon_\mu^\ast \epsilon_\nu^\ast 
 \epsilon^{\mu\nu\alpha\beta} p_\alpha q_\beta\,,
\end{equation}
where $f_P$ is the neutral meson decay constant, $p$ and $q$ are photon momenta, and $\epsilon_\mu$ and $\epsilon_\nu$ are their polarization vectors.
Then the amplitude for the process of interest is given by
\begin{equation}
 i\mathcal{M}\left(P^0 \to \gamma \nu_\alpha \overline{N}\right) = 
 \frac{e^2}{2\pi^2} \frac{d_\gamma^\alpha}{f_P} \epsilon_\mu^\ast \epsilon^{\mu\nu\alpha\beta} 
 \frac{g_{\nu\sigma}}{q^2}
 p_{2\alpha} q_\beta q_\rho 
 \left[\overline{u_\nu} \sigma^{\rho\sigma} P_R v_N\right]\,,
\end{equation}
where $p_2$ is the momentum of the outgoing photon, and $q \equiv p_1 - p_2$ is the momentum of the virtual photon.
Following the conventions of the package~\texttt{FORESEE}~\cite{Kling:2021fwx}, which we will use for our analysis, we express all the scalar products in terms of the angle $\theta$, which is the angle between the direction of the long-lived particle (LLP), $N$ in our case, and the outgoing photon, in the center-of-mass frame of the virtual photon.
Then one has
\begin{equation}
 \left|\mathcal{M}\left(P^0 \to \gamma \nu_\alpha \overline{N}\right)\right|^2 = \frac{e^4}{32\pi^4}
 \frac{\left|d_\gamma^\alpha\right|^2}{f_P^2} 
 \frac{\left(q^2-m_P^2\right)^2 \left(q^2-m_N^2\right)}{q^4} 
 \left[q^2+3m_N^2-\left(q^2-m_N^2\right) \cos2\theta\right]\,.
\end{equation}
Finally, the differential decay rate of interest reads
\begin{equation}
 \frac{\mathrm{d}\Gamma\left(P^0 \to \gamma \nu_\alpha \overline{N}\right)}{\mathrm{d}q^2\,\mathrm{d}\cos\theta} = \frac{1}{512\pi^3 m_P}
 \left(1-\frac{q^2}{m_P^2}\right) 
 \left(1-\frac{m_N^2}{q^2}\right) 
 \left|\mathcal{M}\left(P^0 \to \gamma \nu_\alpha \overline{N}\right)\right|^2\,.
\end{equation}
To get the total decay rate, one needs to integrate over $q^2 \in [m_N^2,m_P^2]$ and $\cos\theta \in [-1,1]$.

The second class of three-body decays is shown in Fig.~\ref{fig:decays}c.
Here, a charged pseudoscalar meson $P^- = \pi^-$, $K^-$, $D^-$, $D_s^-$, $B^-$\footnote{We do not consider $B_c^-$, because their 
production rate at the LHC is negligible with respect to those of the other pseudoscalar mesons~\cite{LHCb:2019tea}.
Also, the inclusion of the charged-conjugated states, $\pi^+, K^+,\ldots$ in the numerical computation is implied.} decays via weak interaction to $\ell_\alpha^-$ and a virtual neutrino, which further converts to a photon and an HNL via $d_\gamma^\alpha$.
The corresponding amplitude is given by
\begin{equation}
 i\mathcal{M}\left(P^- \to \gamma \ell_\alpha^- \overline{N}\right) = i 2\sqrt{2} G_F V_{ij} f_P d_\gamma^\alpha 
 \frac{1}{q^2} k_{1\alpha} \epsilon_\beta^\ast 
 \left[\overline{u_\ell} \slashed{p}_1 \slashed{q} \sigma^{\alpha\beta} P_R v_N\right]\,,
 \label{eq:PgammalN}
\end{equation}
where $G_F = 1.16638\times10^{-5}$~GeV$^{-2}$ is the Fermi constant, $V_{ij}$ is the element of the CKM matrix relevant for a given meson $P^- \sim \overline{u_i} d_j$, and $f_P$ is the decay constant defined through
\begin{equation}
 \langle0|\overline{u_i} \gamma^\mu \gamma_5 d_j|P^-(p_1)\rangle = i f_P p_1^\mu \ ,
\end{equation}
where $p_1$ is the momentum of the meson, $q \equiv p_1-p_2$ is the momentum of the virtual neutrino, and $k_1$ is the momentum of the outgoing photon. 
By expressing the scalar products in terms of the angle $\theta$ between the direction of the LLP and the charged lepton, in the center-of-mass frame of the virtual neutrino, we obtain
\begin{align}
 \left|\mathcal{M}\left(P^- \to \gamma \ell_\alpha^- \overline{N}\right)\right|^2 &= 8 G_F^2 \left|V_{ij}\right|^2 f_P^2 \left|d_\gamma^\alpha\right|^2 
 \frac{\left(q^2-m_N^2\right)^2}{q^4} \bigg[q^2\left(m_P^2-q^2\right)+\left(2q^2+m_P^2\right)m_{\ell_\alpha}^2 \nonumber \\
 &\phantom{{}={}}-m_{\ell_\alpha}^4+\left(q^2-m_{\ell_\alpha}^2\right)\sqrt{\lambda\left(q^2,m_P^2,m_{\ell_\alpha}^2\right)}\cos\theta\bigg]\,,
\end{align}
where $\lambda(x,y,z)\equiv(x-y-z)^2-4yz$ is the K\"all\'en function.
Finally, the corresponding differential decay rate yields
\begin{equation}
 \frac{\mathrm{d}\Gamma\left(P^- \to \gamma \ell_\alpha^- \overline{N}\right)}{\mathrm{d}q^2\,\mathrm{d}\cos\theta} = \frac{\sqrt{\lambda\left(q^2,m_P^2,m_{\ell_\alpha}^2\right)}}{512\pi^3 m_P^3} \left(1-\frac{m_N^2}{q^2}\right) 
 \left|\mathcal{M}\left(P^- \to \gamma \ell_\alpha^- \overline{N}\right)\right|^2\,.
\end{equation}
The total decay rate can be obtained by integrating over $q^2 \in [m_N^2,(m_P-m_{\ell_\alpha})^2]$ and $\cos\theta \in [-1,1]$.

\begin{figure}[t]
 \hspace{1cm}
 \includegraphics[width=0.85\textwidth]{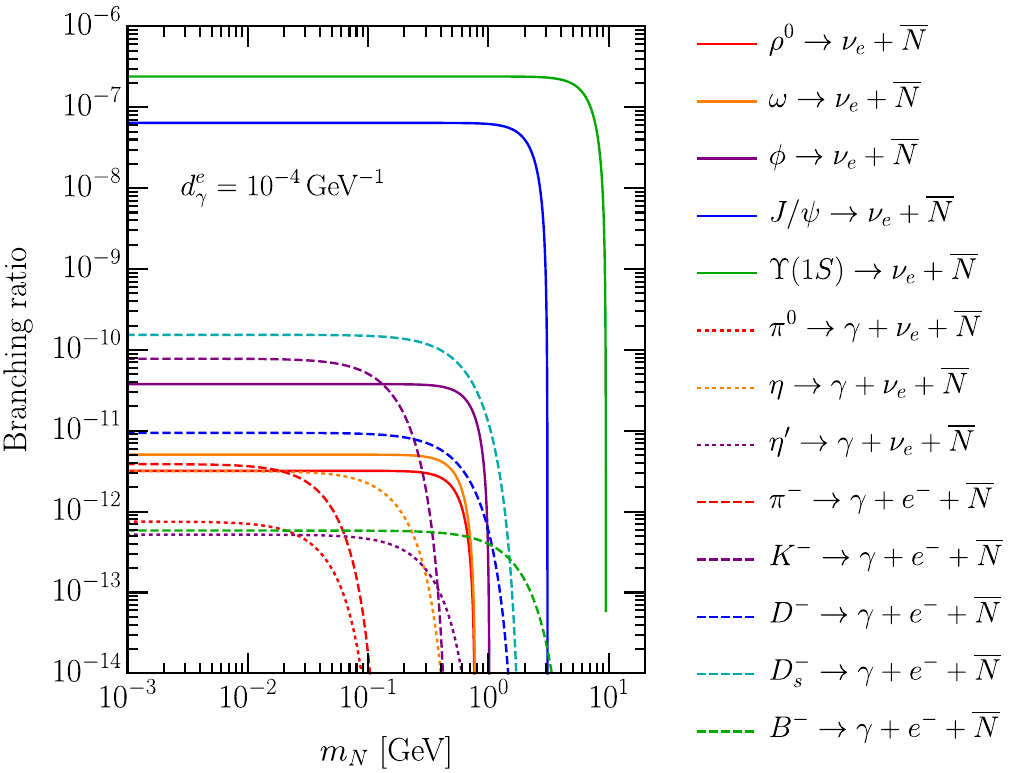}\\[0.5cm]
 \hspace*{1cm}
 \includegraphics[width=0.85\textwidth]{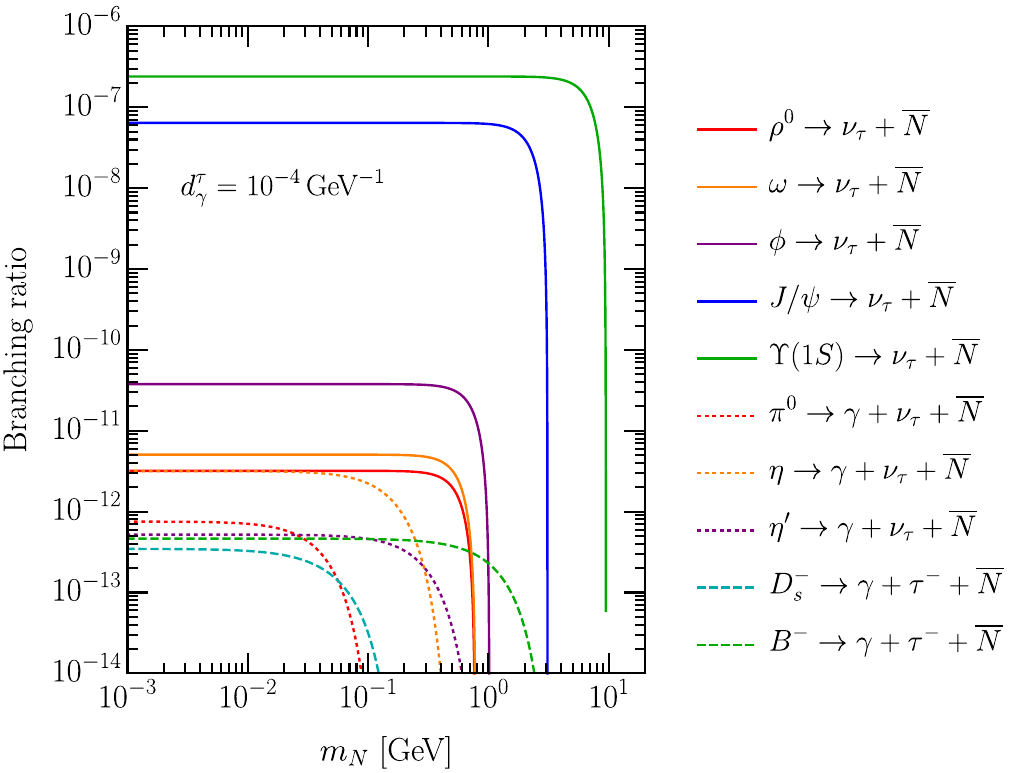} 
 \caption{Branching ratios of meson decays mediated by the neutrino dipole operator $d_\gamma^\alpha$ of electron flavor (top) and of tau flavor (bottom).}
 \label{fig:BRs}
\end{figure}

In Fig.~\ref{fig:BRs}, we plot the branching ratios of the two- and three-body decays discussed above.
In the upper plot, we assume an exclusive dipole coupling to the electron neutrino, with value $d_\gamma^e = 10^{-4}$~GeV$^{-1}$, whereas in the lower plot, we take the same value for an exclusive coupling to the $\tau$ neutrino.
The results for the muon flavor are very similar to those shown for the electron flavor in the upper plot, with the only difference being due to phase space effects. 
They are relevant only for the branching ratio of $\pi^- \to \gamma \mu^- \overline{N}$, which is smaller than that of $\pi^- \to \gamma e^- \overline{N}$ because of the similar masses of $\pi^-$ and $\mu^-$.
As can be inferred from Eq.~\eqref{eq:GammaV0nuNDir}, the two-body decay width is proportional to the mass of the parent vector meson. 
Together with the larger values of the decay constant $f_V$ (see Tab.~\ref{tab:decayconstants}), this explains why $\mathrm{BR}(\Upsilon(1S)\to\nu_\alpha \overline{N})$ and $\mathrm{BR}(J/\psi \to \nu_\alpha \overline{N})$ are more than three orders of magnitude larger than the branching ratios of lighter meson decays.
Some of the three-body decays, in particular those of $K^-$, $D^-$ and $D_s^-$ in the case of $d_\gamma^e \neq 0$, have larger branching ratios than the two-body decays of light vector mesons $\rho^0$ and $\omega$.

Finally, we provide in Tab.~\ref{tab:meson_number} a comprehensive list of the production cross sections of the considered mesons for the forward hemisphere at the LHC, extracted from \texttt{FORESEE}.
For the charged mesons, we list the production cross sections only for one of the two charged-conjugated states.

\begin{table}[t]
	\centering
	\begin{tabular}{|l|c||l|c||l|c|}
        \hline
		Meson 
        & Cross section~[fb]   &  Meson 
        & Cross section~[fb] & 
        Meson 
        & Cross section~[fb] \\
        \hline
		$\rho^0$ 
        & $1.86 \times 10^{14}$ & 
        $\pi^0$ 
        & $1.54 \times 10^{15}$ & 
        $\pi^\pm$ 
        &$1.35 \times 10^{15}$ \\
        \hline
  		$\omega$ 
        &$1.74 \times 10^{14}$ & 
        $\eta$ 
        & $1.69 \times 10^{14}$ & 
        $K^\pm$ 
        & $1.57 \times 10^{14}$ \\
        \hline
    	$\phi$ 
        & $2.09 \times 10^{13}$ & 
        $\eta^\prime$ 
        & $1.81 \times 10^{13}$ & 
        $D^\pm$ 
        & $8.48 \times 10^{11}$ \\ 
        \hline
        $J/\psi$ 
        & $3.93 \times 10^{10}$  & 
        & & 
        $D_s^\pm$ 
        & $2.70 \times 10^{11}$ \\ 
        \hline
		$\Upsilon(1S)$ 
        & $5.30 \times 10^{8}$ & 
        & & 
        $B^\pm$ 
        &$8.23 \times 10^{10}$ \\ 
        \hline
	\end{tabular}
	\caption{Production cross sections of mesons at the 14-TeV LHC in the forward hemisphere~\cite{Kling:2021fwx}.
    For the charged mesons, the given cross sections are only for one of the two charge-conjugated states.
 }
	\label{tab:meson_number}
\end{table}

\subsection{HNL decay}\label{subsec:HNL_decay}

The dipole interaction of Eq.~\eqref{eq:dipole_IR} would additionally result in the HNL decay into a neutrino and a photon. 
For a Dirac HNL, the corresponding decay width is given by~\cite{Magill:2018jla}
\begin{equation}
 \Gamma\left(N \to \nu_\alpha \gamma\right) = \frac{\left|d_\gamma^\alpha\right|^2 m_N^3}{4\pi}\,.
\end{equation}
For HNL energies $E_N \gg m_N$, the product $\beta\gamma \approx E_N/m_N$, and the HNL decay length in the laboratory frame reads
\begin{equation}\label{eq:decay_length}
 \ell_\mathrm{dec} = \beta\gamma c \tau \approx \frac{4\pi c E_N}{\left|d_\gamma^\alpha\right|^2 m_N^4}\,.
\end{equation}
In particular, assuming $E_N \approx 100$~GeV, which is a typical energy of an HNL produced in meson decays at the 14-TeV LHC,
we find
\begin{equation}
 \ell_\mathrm{dec} \approx 490~\text{m}~ \left(\frac{10^{-6}~\text{GeV}^{-1}}{d_\gamma^\alpha}\right)^2 \left(\frac{0.15~\text{GeV}}{m_N}\right)^4\,,
\end{equation}
which is around the length scale aimed to be probed by the LLP detectors such as FASER and FACET.


\section{Experiments \& simulation}\label{sec:exp}

HNLs with masses in the MeV to GeV range are generically expected to be long-lived, when interacting with the SM fields only through the active-sterile mixing.
This is also true in the presence of higher-dimensional operators, as the one considered in this work, if the cut-off scale is around the TeV or above as shown in Eq.~\eqref{eq:decay_length}, see also \textit{e.g.} Ref.~\cite{Barducci:2022hll}.
This behavior leads to exotic signatures such as displaced vertices that might be detected at a macroscopic distance away from the IP.
Multiple dedicated concepts of ``far detectors'' within the LHC facility have been proposed to detect HNLs among other LLP candidates, such as 
AL3X~\cite{Gligorov:2018vkc}, ANUBIS~\cite{Bauer:2019vqk}, CODEX-b~\cite{Gligorov:2017nwh}, FACET~\cite{Cerci:2021nlb}, FASER~\cite{FASER:2018eoc}, MATHUSLA~\cite{Curtin:2018mvb}, and MoEDAL-MAPP~\cite{Pinfold:2019nqj, Acharya:2022nik}. For a recent review on the status of HNLs searches see \textit{e.g.}~Ref.~\cite{Abdullahi:2022jlv}.
Among these proposals, FASER and FACET are the most promising candidates for discovering light LLPs produced from meson decays, in virtue of their very forward location with respect to the ATLAS and CMS IPs, respectively.
We hence choose to concentrate on these two experiments in this work.

\paragraph{FASER} is located 480 meters away from the IP of the ATLAS detector, in the very forward direction, with a polar angle with respect to the beam direction $\theta \lesssim 10^{-3}$.
It has been installed and is under operation during Run~3 of the LHC, having already delivered its first physics results~\cite{Petersen:2023hgm,FASER:2023tle}.
Given the relatively small solid-angle coverage of FASER and the relatively limited data set ($\approx150$~fb$^{-1}$ integrated luminosity by the end of Run~3) at disposal, we find that FASER is expected to observe too few signal events for the scenario under consideration to be statistically meaningful.

However, FASER is planned to have a follow-up upgraded program known as FASER2 to be running during the high-luminosity LHC (HL-LHC) era.
The geometrical information of FASER2 is summarized below,
\begin{gather}
	L_{xy} < 1~\text{m}\,,
 \qquad
	 475~\text{m} < L_z < 480~\text{m} \,,
\end{gather} 
where $L_{xy}$ and $L_z$ represent the distance to the IP in the transverse and longitudinal direction,\footnote{It is also likely that FASER2 would be constructed in the proposed Forward Physics Facility~\cite{Feng:2022inv}, which is located 620 m away from the ATLAS IP in the very forward direction.
However, we expect the difference in the longitudinal distance between 480 m and 620 m would only lead to negligible discrepancies in the final sensitivity of our results.} respectively, and the integrated luminosity of FASER2 will be 3000 fb$^{-1}$.
With a larger volume and more data collected, FASER2 should have stronger sensitivities than FASER.
In this work, we will consider FASER2 for the sensitivity study.

\paragraph{FACET} is another recently proposed far-detector program at the LHC.
Being a sub-system of the CMS experiment, it is supposed to be enclosing the beam pipe, covering the polar-angle range between 1 and 4 mrad.
Compared to FASER, it is relatively close to the corresponding IP, with a distance slightly over 100 m.
It will have a length of 18\;m and an integrated luminosity of 3000 fb$^{-1}$.
The geometrical information is
\begin{gather}
	0.18~\text{m} < L_{xy} < 0.5~\text{m}\,,
 \qquad 
	 101~\text{m} < L_z < 119~\text{m} \,.
\end{gather}

In order to estimate the signal-event rates at FASER2 and FACET, for decays of HNLs produced from rare meson decays, we utilize the publicly available Python-based package \texttt{FORESEE}~\cite{Kling:2021fwx}.
The package has already implemented the kinematic distributions in one forward hemisphere for the various types of mesons at the LHC with the center-of-mass energy 14 TeV that we consider in this work.
Further, once we provide the formulae for the decay branching ratios of the mesons into the HNL $N$, \texttt{FORESEE} automatically computes the spectra $\mathrm{d}\sigma_N^M/(\mathrm{d}p_N\, \mathrm{d}\cos\theta_N)$ of the HNLs produced from meson $M$ in the kinematic space spanned by the 3-momentum magnitude $p_N$ and polar angle $\theta_N$.
We thus obtain the signal-event rates $N_S$ as
\begin{equation}
 N_S =   \mathcal{L}_{\text{int}}   \sum_{M} \int \mathrm{d}p_N\, \mathrm{d}\cos\theta_N     \frac{\mathrm{d}\sigma_N^M}{\mathrm{d}p_N\, \mathrm{d}\cos\theta_N}  P(p_N, \cos\theta_N, c\tau_N)\cdot \epsilon_{\text{kin}}\cdot\text{BR}(N \rightarrow \nu \gamma)\,,
\end{equation}
where $\mathcal{L}_{\text{int}}=3000$ fb$^{-1}$ is the integrated luminosity of either FASER2 or FACET, the summation goes over all the contributing mesons' types, $\text{BR}(N \rightarrow \nu \gamma)$ is the decay branching ratio of the HNL into the signal final state, and $\epsilon_{\text{kin}}$ reflects the kinematical efficiency from the cut on the HNL momentum, $p_N > 100$~GeV, which is implemented in \texttt{FORESEE} by default (for FASER and FASER2) and is assumed to be able to cut away background events~\cite{Kling:2021fwx}.
We further assume it is the case also for FACET.
$P(p_N,\cos\theta_N,c\tau_N)$ is the decay probability of the HNL inside the fiducial volume of the considered experiments, estimated with the following exponential distribution law:
\begin{equation}
 P(p_N,\cos\theta_N,c\tau_N) =
 \begin{cases}
 e^{-\frac{L_1 m_N}{p_N \cos\theta_N\, c \tau_N }}  -  e^{-\frac{L_2 m_N}{p_N \cos\theta_N\, c \tau_N}}\,,\quad\text{if hitting detector,}\\
 0\,, \quad \text{if missing detector,}
 \end{cases}
\end{equation}
where $L_1$ and $L_2$ are the longitudinal distances from the IP to the near and far ends of the forward detectors, respectively, and $c\tau_N$ is the proper decay length of the HNL, see Sec.~\ref{subsec:HNL_decay}.

Before moving to the presentation of the numerical results of the experimental sensitivities, a comment regarding possible sources of backgrounds for the signal process of our interest, which is a highly energetic single photon, is in order.
We first discuss the background events at the FASER2 experiment.
They can arise from \textit{e.g.} neutrino interaction via SM CC interactions deep inside the calorimeter, which can be identified by the pre-shower station~\cite{FASER:2021cpr,Boyd:2803084} installed just before the calorimeter itself. 
Other background events may stem from neutral pion decays into a pair of photons where only one photon is identified, and from high-energy neutrinos and muons produced at the IP and reaching the detectors leading to interactions with the detector.
These events can, in principle, all be eliminated with different approaches.
For more detail, we refer the reader to Refs.~\cite{Jodlowski:2020vhr,FASER:2018bac,Dreiner:2022swd,Dienes:2023uve}.
We therefore argue that the working assumption of vanishing background for FASER2 is legitimate, as usually assumed in phenomenological studies.
However, for FACET, the same arguments may not apply, since the experiment is located at the beam-pipe area.
In this case, one should be more careful with the estimation of possible background sources.
Nevertheless, for simplicity, we choose to be optimistic and assume that new experimental approaches will be proposed to essentially achieve a zero background rate, without reducing the signal-event rates.
Admittedly, a full simulation would be required in order to know the exact number of such background events at either FASER2 or FACET.
However, such a simulation is beyond the scope of this work, which therefore presents the optimal reach of these facilities.
To summarize, we assume negligible background rates for both FASER2 and FACET, and hence show the sensitivity reach of both experiments at 95\% confidence level (CL) by requiring a number of signal event $N_S=3$.


\section{Numerical results}\label{sec:results}

\begin{figure}[t]
\centering
\includegraphics[width=0.9\textwidth]{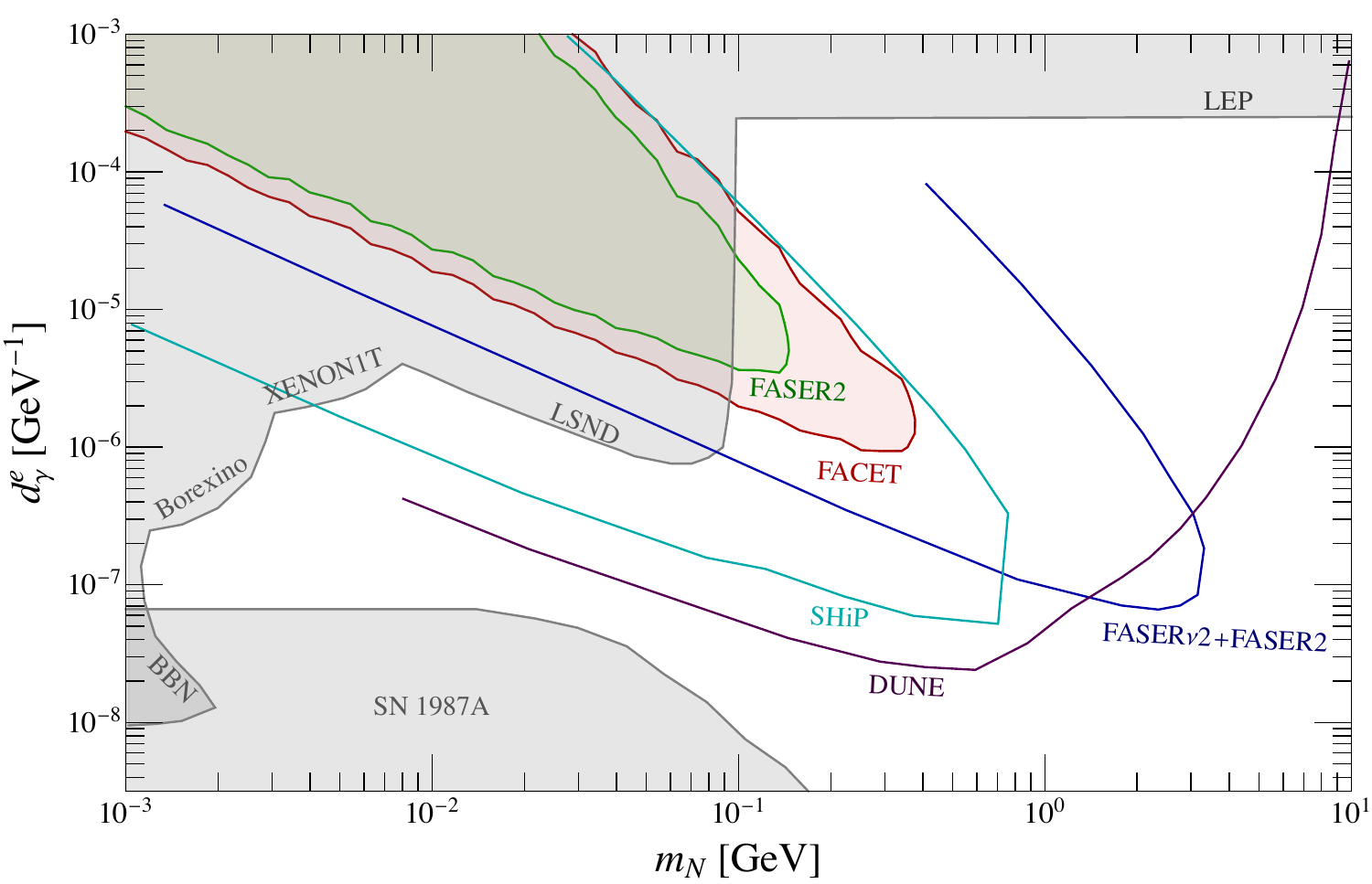}
\caption{The 95\% CL~sensitivity to $d_{\gamma}^e$ as a function of $m_N$ with the search for single photons in the decay vessel of the FASER2~(shaded green) and FACET~(shaded red) detectors. 
The photons are assumed to originate from HNLs produced in meson decays at the HL-LHC.
The current best limits from XENON1T~\cite{Brdar:2020quo}, Borexino~\cite{Brdar:2020quo}, SN 1987A~\cite{Magill:2018jla}, and LSND~\cite{Magill:2018jla} are shown in the gray shaded region.
Note that the limits from LEP are very sensitive to the model parameter $a$~\cite{Zhang:2023nxy}.
Here we take $a=2\tan\theta_w$ for LEP as a sample benchmark. 
For comparison purpose, the expected reach from the secondary production of HNLs in neutrino scatterings in the FASER$\nu$2 detector as well as the subsequent search for single photons in the decay vessel of FASER2 is also shown (blue line, labelled as ``FASER$\nu$2+FASER2"), reproduced from Ref.~\cite{Jodlowski:2020vhr}. 
The expected sensitivities of the proposed SHiP~\cite{Magill:2018jla} and DUNE~\cite{Ovchynnikov:2022rqj} experiments are represented by the cyan and purple lines, respectively.
The DUNE sensitivity is a combined result of the mono-photon, double-bang, and di-lepton signatures.
See the text for further details.}
\label{fig:fdgA}
\end{figure}

\begin{figure}
\centering
\includegraphics[width=0.9\textwidth]{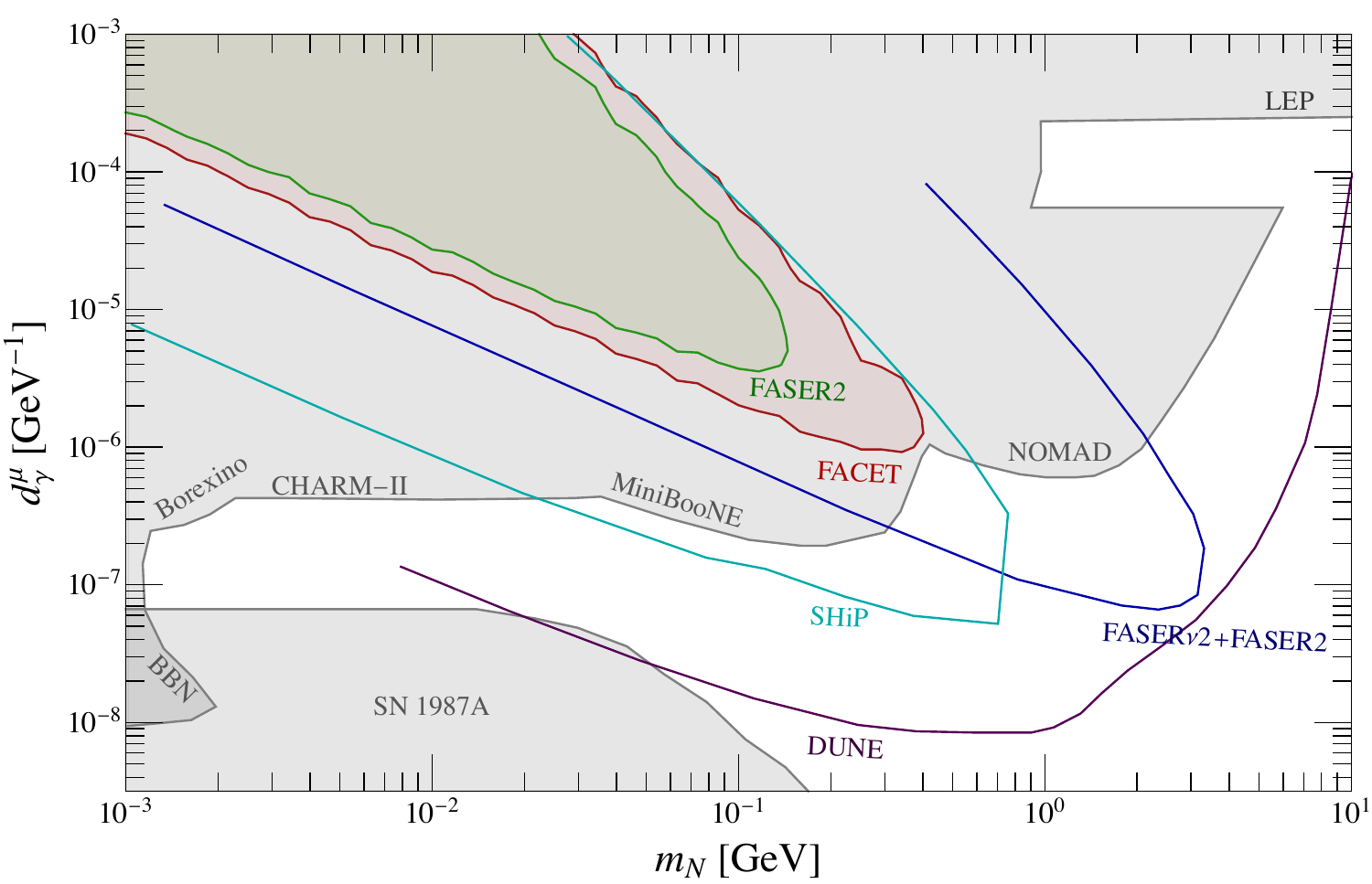}
\caption{The same as in Fig.~\ref{fig:fdgA}, but for $d_{\gamma}^{\mu}$. 
Current best limits are extracted from Borexino~\cite{Brdar:2020quo}, SN 1987A~\cite{Magill:2018jla}, CHARM-II~\cite{Coloma:2017ppo}, MiniBooNE~\cite{Magill:2018jla}, and NOMAD~\cite{Magill:2018jla}.
}
\label{fig:fdgB}
\end{figure}

\begin{figure}
\centering
\includegraphics[width=0.9\textwidth]{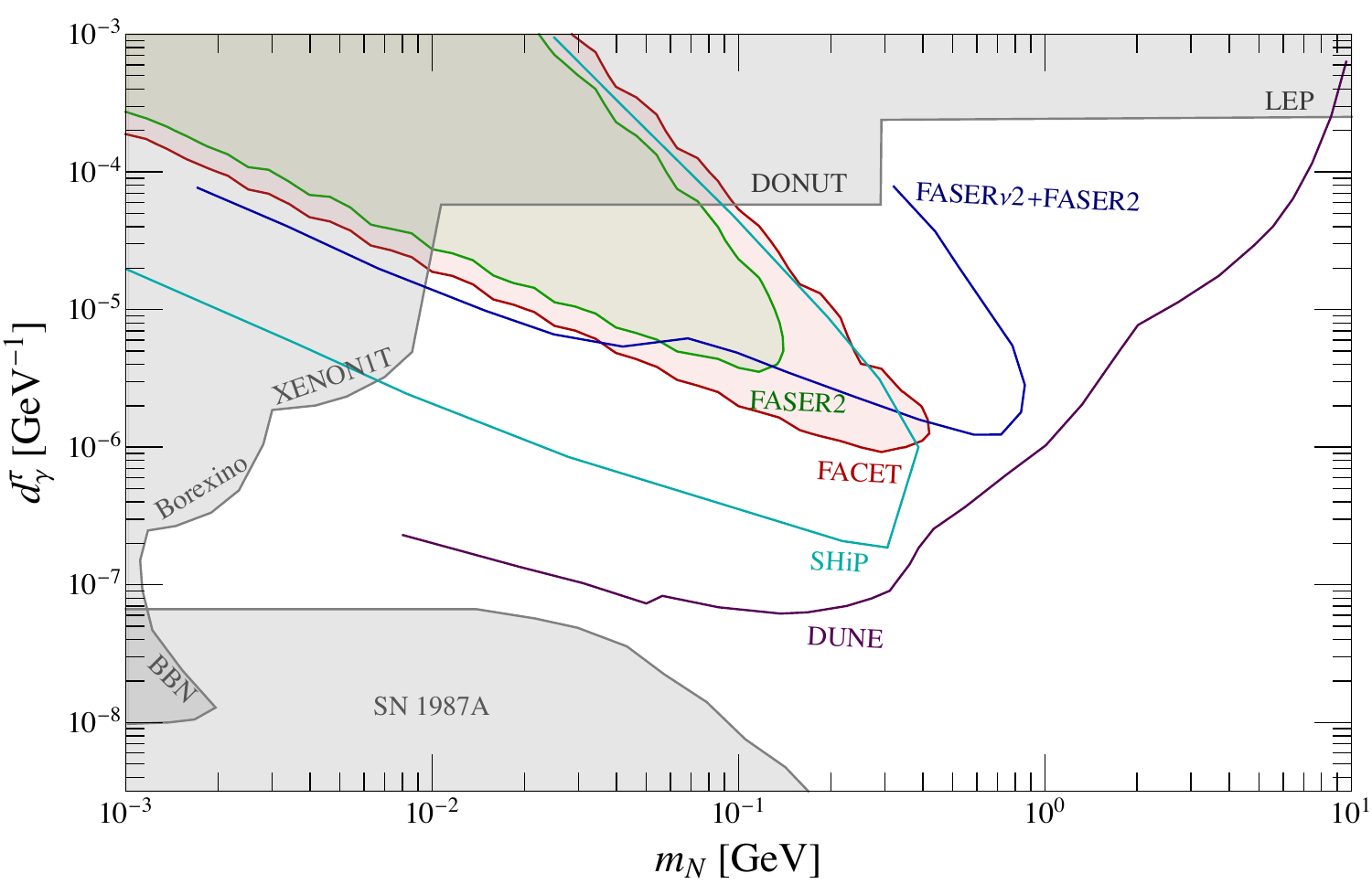}
\caption{The same as in Fig.~\ref{fig:fdgA}, but for $d_{\gamma}^{\tau}$.
Current best limits are extracted from XENON1T~\cite{Brdar:2020quo}, Borexino~\cite{Brdar:2020quo}, SN 1987A~\cite{Magill:2018jla}, and DONUT~\cite{DONUT:2001zvi}.
}
\label{fig:fdgC}
\end{figure}

In this section, we present the numerical results for the sensitivity of the far-detector experiments to the dipole-portal couplings to sterile neutrino, produced from meson decays at the HL-LHC. 
We show our main findings in Fig.~\ref{fig:fdgA}, Fig.~\ref{fig:fdgB}, and Fig.~\ref{fig:fdgC} for the cases of dipole-portal coupling to electron, muon, and tau neutrinos, respectively.
In particular, the parts of the parameter space to which the future FASER2 and FACET experiments are sensitive at 95\% CL~are shaded in green and red, respectively.

In the figures we also show in gray the current constraints arising from terrestrial and astrophysical searches. 
These include limits from Borexino~\cite{Brdar:2020quo}, XENON1T~\cite{Brdar:2020quo}, CHARM-II~\cite{Coloma:2017ppo}, MiniBooNE~\cite{Magill:2018jla}, LSND~\cite{Magill:2018jla}, NOMAD~\cite{Gninenko:1998nn,Magill:2018jla}, DONUT~\cite{DONUT:2001zvi} and LEP~\cite{Zhang:2023nxy}, which are all terrestrial experiments.   
The gray shaded region also includes cosmological and astrophysical constraints from the Big Bang Nucleosynthesis (BBN) data and observations of Supernova SN 1987A~\cite{Magill:2018jla}, which become important for light ($m_N \lesssim 100$~MeV) and very weakly coupled sterile neutrinos.
Note that for the constraints from XENON1T, Borexino, and SN 1987A, lepton-flavor-universal couplings were assumed. 
These existing constraints as well as the results obtained in this work do not depend on the ratio $a = d_{\calW}^\alpha/d_{\cal B}^\alpha$, see Eq.~\eqref{eq:dzdw}, since the typical scattering energies are far below the EW scale, while for LEP, the limits are very sensitive to the parameter $a$.
The detailed discussion for LEP can be found in Ref.~\cite{Zhang:2023nxy}, while here we fix $a=2\tan\theta_w$ as a benchmark for LEP for illustrative purpose. 
For this value of $a$, the dipole coupling to the $Z$ boson, $d_Z^\alpha$, vanishes, see Eq.~\eqref{eq:dzdw}.

Considering the high-energy neutrinos produced at the LHC and the HL-LHC, Ref.~\cite{Jodlowski:2020vhr} investigated the sensitivity reach on the dipole-portal coupling to the sterile neutrinos at FASER and its possible successor FASER2, along with their neutrino sub-detectors FASER$\nu$ and FASER$\nu$2.
In Ref.~\cite{Jodlowski:2020vhr}, the authors primarily focus on the secondary production of HNLs from neutrino scatterings in the close vicinity of, or inside, the FASER$\nu$2 detector, as well as on subsequent detectable signals of a single photon arising from the decay $N\to\nu\gamma$ in the decay vessel of FASER2. 
We reproduce their results with blue lines (which are labelled as ``FASER$\nu$2+FASER2" in the figures) for comparison purpose,\footnote{We note however that Ref.~\cite{Jodlowski:2020vhr} assumes a universal dipole coupling to all neutrino flavors, while we work under the assumption of one flavor dominance \textit{e.g.} $d_\gamma^e \neq 0$ and $d_\gamma^\mu = d_\gamma^\tau = 0$.} since they investigate the same signature in the same detector as us.
Besides, we also show the expected reach of the proposed SHiP (cyan lines) and DUNE (purple lines) experiments following Ref.~\cite{Magill:2018jla} and Ref.~\cite{Ovchynnikov:2022rqj}, respectively. 
In the case of DUNE, the resulting sensitivity comes from several experimental signatures at either near detector (for $d_\gamma^e$ and $d_\gamma^\mu$) or both near and far detectors (for $d_\gamma^\tau$).
These signatures include mono-photon events, double-bang events, and di-lepton events. 
The latter are induced by the three-body HNL decays $N \to \nu_\alpha \ell^+ \ell^-$, $\ell = e$, $\mu$, mediated by a virtual photon, see Ref.~\cite{Ovchynnikov:2022rqj} for more details.

For the numerical results of the $\nu_e$-specific dipole-portal coupling to HNLs, $d_\gamma^e$, given in Fig.~\ref{fig:fdgA}, one finds that the flavor-universal constraints from XENON1T~\cite{Brdar:2020quo}, Borexino~\cite{Brdar:2020quo}, and SN 1987A~\cite{Magill:2018jla}, along with the constraints from LSND~\cite{Magill:2018jla}, exhibit the current best limits for $m_N \lesssim 0.1$ GeV.
FASER2 can probe unexplored parameter space for the HNL mass approximately between 0.1 GeV and 0.15 GeV.
In detail, the upper limits on $d_\gamma^e$ can be down to about $3.5 \times 10^{-6}$ GeV$^{-1}$ for $m_N \approx 0.14$ GeV.
Thanks to the larger solid-angle coverage and a bigger length, FACET can be sensitive to an even larger parameter region than FASER2.
Relative to the existing bounds, FACET can provide competitive sensitivities to $d_\gamma^e$ for $0.1\text{ GeV} \lesssim m_N \lesssim 0.4$ GeV, where the upper limits on $d^e_\gamma$ can be as low as about $9.4 \times 10^{-7}$~GeV$^{-1}$ for $m_N$ around 0.3 GeV.
Finally, we note that Ref.~\cite{Jodlowski:2020vhr} shows that combining FASER$\nu$2 and FASER2 allows to even further improve the expected sensitivities.

Regarding the $\nu_\mu$-specific dipole coupling to HNLs, $d_\gamma^\mu$, considered in Fig.~\ref{fig:fdgB}, the existing constraints can be obtained from various experiments~\cite{Magill:2018jla}.
The best current limits come from the combination of CHARM-II~\cite{Coloma:2017ppo}, MiniBooNE~\cite{Magill:2018jla}, and NOMAD~\cite{Magill:2018jla} in the mass range of $3 \text{ MeV} - 6$ GeV. 
Unfortunately, one observes that neither FASER2 nor FACET can compete with the neutrino experiment MiniBooNE~\cite{Magill:2018jla}.

Finally, Fig.~\ref{fig:fdgC} shows the numerical results where the HNLs are coupled via the dipole-portal coupling to the tau neutrino, $d_\gamma^\tau$.
Here, for the strongest existing bounds, besides the flavor-universal constraints from XENON1T~\cite{Brdar:2020quo}, Borexino~\cite{Brdar:2020quo}, and SN 1987A~\cite{Magill:2018jla}, we only find as relevant the upper limits from DONUT~\cite{DONUT:2001zvi} at 90\% CL~of $5.8 \times 10^{-5}$ GeV$^{-1}$ for $0.01 \lesssim m_N \lesssim  0.3$ GeV.
Via meson decays at the HL-LHC, FASER2 can improve the constraints on $d_\gamma^\tau$ down to about $3.5 \times 10^{-6}$~GeV$^{-1}$ for $m_N\lesssim0.12$ GeV, which is more than one order of magnitude stronger than the bound from DONUT.
Further, for $m_N \approx 0.1$~GeV, displaced decays of the HNLs produced from meson decays can probe smaller values of $d^\tau_\gamma$ than the secondary production in neutrino scattering at FASER$\nu$2~\cite{Jodlowski:2020vhr}.
Finally, FACET shows even stronger expected sensitivity reach.
It outperforms the combination of FASER$\nu$2 and FASER2~\cite{Jodlowski:2020vhr} for the $m_N$ range of $0.04\text{ GeV} - 0.4$~GeV.

Before we close the section, we comment on the relative importance of the contributions from two-body and three-body decays of the mesons to the signal-event rates in the three lepton-flavor cases.
The production rates of the HNLs from either the two-body or three-body decays of the \textit{neutral} mesons are flavor-independent, since the final-state active neutrinos are all (almost) massless.
However, in the case of three-body decays of the \textit{charged} mesons, the HNL production rates are flavor-dependent as a result of the different masses of the final-state charged leptons, \textit{cf}.~Fig.~\ref{fig:BRs}.
Nevertheless, although the production rates of the HNLs from the neutral and charged mesons are comparable, the contributions from the three-body decays will be very strongly suppressed once we impose the requirement on the HNL three-momentum magnitude, $p_N>100$ GeV.
This is because, compared to the two-body decays, the three-body decays of the mesons lead to HNLs with smaller $p_N$ and larger polar angles.
As a result, across Figs.~\ref{fig:fdgA}--\ref{fig:fdgC} for the HNLs coupled to neutrinos of different lepton flavors, the obtained sensitivity results are similar.


\section{Conclusions}\label{sec:conclusions}

In this work, we have studied Dirac-type sterile neutrinos, also known as heavy neutral leptons (HNLs), coupled to active neutrinos via a dipole operator with the standard-model photon.
We have focused on light, (sub-)GeV-scale HNLs produced from the rare decays of QCD mesons, 
which are copiously produced at the high-luminosity LHC (HL-LHC).
We have worked out in detail the decay branching ratios of various pseudoscalar and vector mesons into the HNL, for both two-body or three-body decays.
Such HNLs are generally expected to be long-lived, leading to exotic signatures of displaced objects, as a single photon from 
the HNL decay into an active neutrino and a photon considered in this work.

Proceeding to the numerical investigation, we have considered the proposed forward far-detector programs, FASER2 and FACET at the HL-LHC, and, utilizing the Python-based tool \texttt{FORESEE}, evaluated their sensitivities to light, long-lived HNLs in the parameter space spanned by the HNL mass $m_N$ and the strength of the dominant dipole interaction $d_\gamma^\alpha$ with $\alpha=e, \mu, \tau$, separately.
Generally, we observe the dominant contributions to the signal-event rates to stem from the two-body decays of flavorless vector mesons, mainly as a result of the requirement that the three-momentum magnitude of the HNL should be larger than 100 GeV, which is usually assumed to be able to reduce the background to a negligible level.
We thus assumed vanishing background for both FASER2 and FACET,\footnote{As discussed in the main text, this should be a good assumption for FASER2, while for FACET more sophisticated studies are required.} and have obtained the exclusion bounds at 95\% confidence level by requiring three signal events.
Taking into account the best current limits, we find that in the cases of the HNLs coupled to the electron and tau neutrinos, FASER2 and FACET can probe relatively large unexplored parts of the parameter space, while for the muon case all the sensitive parameter regions have been already ruled out.
Finally, we note that FACET in general outperforms FASER2, thanks to its closer location to the IP, a larger solid-angle coverage, as well as its larger length, although this holds only in the limit of zero background.

To conclude, we find that the proposed experiments FASER2 and FACET can improve the current limits on the dipole portal couplings $d^e_\gamma$ and $d_\gamma^\tau$ by up to about two orders of magnitude for $m_N\lesssim 0.4$ GeV. 
The considered physics case further motivates the construction and operation of these future experiments in the HL-LHC era.

\section*{Acknowledgements}


WL is supported by National Natural Science Foundation of China (Grant No.12205153), and the 2021 Jiangsu Shuangchuang (Mass Innovation and Entrepreneurship) Talent Program (JSSCBS20210213).
YZ is supported  by the National Natural Science Foundation of China (Grant No. 11805001) and the Fundamental Research Funds for the Central Universities (Grant No. JZ2023HGTB0222).

\appendix

\section{Meson masses, decay widths, and decay constants}
\label{sec:appendix}

In Tab.~\ref{tab:decayconstants}, we provide the masses, the decay widths, and the decay constants of vector and pseudoscalar mesons employed in this work.
\begin{table}[t]
 \renewcommand{\arraystretch}{1.2}
 \centering
 \begin{tabular}{| l | l | l | l |}
 \hline
 Meson & Mass $m_{V/P}$~[MeV] & 
 Total decay width~[MeV] & 
 Decay constant $f_{V/P}$~[MeV]\\
 \hline
 $\rho^0$ & 775.26 (0.23) & 147.4 (0.8) & 216 (3)~\cite{Bharucha:2015bzk} \\ 
 $\omega$ & 782.66 (0.13) & 8.68 (0.13) & 197 (8)~\cite{Bharucha:2015bzk} \\
 $\phi$ & 1019.461 (0.016) & 4.249 (0.013) & 233 (4)~\cite{Bharucha:2015bzk} \\
 $J/\psi$ & 3096.900 (0.006) & 0.0926 (0.0017) & 407 (5)~\cite{Donald:2012ga} \\
 $\Upsilon(1S)$ & 9460.30 (0.26) & 0.05402 (0.00125) & 689 (5)~\cite{Colquhoun:2014ica} \\
 \hline
 $\pi^0$ & 134.9768 (0.0005) & $7.80\times10^{-6}$ & 92.1 (0.6)~\cite{FlavourLatticeAveragingGroupFLAG:2021npn} \\ 
 $\eta$ & 547.862 (0.017) & $1.31~(0.05)\times10^{-3}$ & 115.0 (2.8)~\cite{Bali:2021qem} \\  
 $\eta^\prime$ & 957.78 (0.06) & 0.188 (0.006) & 100.1 (3.0)~\cite{Bali:2021qem} \\  
 \hline
 $\pi^\pm$ & 139.57039 (0.00018) & $2.53\times10^{-14}$ & 130.2 (0.8)~\cite{FlavourLatticeAveragingGroupFLAG:2021npn} \\
 $K^\pm$ & 493.677 (0.016) & $5.31\times10^{-14}$ & 155.7 (0.3)~\cite{FlavourLatticeAveragingGroupFLAG:2021npn} \\
 $D^\pm$ & 1869.66 (0.05) & $6.33\times10^{-10}$ & 212.0 (0.7)~\cite{FlavourLatticeAveragingGroupFLAG:2021npn} \\
 $D_s^\pm$ & 1968.35 (0.07) & $1.31\times10^{-9}$ & 249.9 (0.5)~\cite{FlavourLatticeAveragingGroupFLAG:2021npn} \\
 $B^\pm$ & 5279.34 (0.12) & $4.02\times10^{-10}$ & 190.0 (1.3)~\cite{FlavourLatticeAveragingGroupFLAG:2021npn} \\
 \hline
 \end{tabular}
 \caption{Masses and total decay widths~\cite{ParticleDataGroup:2022pth} along with decay constants of vector and pseudoscalar mesons employed in this work. 
 For $V^0=\rho^0$, $\omega$, $\phi$, $J/\psi$, $\Upsilon(1S)$, we use the values extracted from the experimentally measured $\Gamma(V^0 \to e^+ e^-)$ 
 as explained in Refs.~\cite{Bharucha:2015bzk,Donald:2012ga,Colquhoun:2014ica}.}
 \label{tab:decayconstants}
\end{table}
In addition, to compute the three-body decays in Eq.~\eqref{eq:PgammalN}, we have used the values of the CKM matrix elements $|V_{ij}|$ given in Ref.~\cite{UTfit:2022hsi}.

\bibliographystyle{JHEP}
\bibliography{dipole}

\end{document}